# A Generalizable Artificial Intelligence Model for COVID-19 Classification Task Using Chest X-ray Radiographs: Evaluated Over Four Clinical Datasets with 15,097 Patients

Ran Zhang[1] Ph.D., Xin Tie[1] MS, John W. Garrett[3,1] Ph.D., Dalton Griner[1] MS, Zhihua Qi[2] Ph.D., Nicholas B. Bevins[2] Ph.D., Scott B. Reeder[3,1,4,5,6] MD/Ph.D., and Guang-Hong Chen[1,3] Ph.D.

1. Department of Medical Physics, School of Medicine and Public Health, University of Wisconsin, Madison, WI 53705, USA

2. Department of Radiology, Henry Ford Health, Detroit, MI 48202, USA

3. Department of Radiology, School of Medicine and Public Health, University of Wisconsin, Madison, WI 53792, USA

4. Department of Biomedical Engineering, University of Wisconsin, 1550 Engineering Dr, Madison, WI, 53706, USA

5. Department of Medicine, University of Wisconsin, 1685 Highland Ave, Madison, WI, 53792, USA

6. Department of Emergency Medicine, University of Wisconsin, 800 University Bay Dr Suite 310, Madison, WI, 53705, USA

**Address correspondence to:**

Guang-Hong Chen, Ph.D.,

Department of Medical Physics and Department of Radiology,

School of Medicine and Public Health,

University of Wisconsin in Madison, Madison, WI 53705,

Email: gchen7@wisc.edu

# Abstract

**Purpose**

To answer the long-standing question of whether a model trained from a single clinical site can be generalized to external sites.

**Materials and Methods**

17,537 chest x-ray radiographs (CXRs) from 3,264 COVID-19-positive patients and 4,802 COVID-19-negative patients were collected from a single site for AI model development. The generalizability of the trained model was retrospectively evaluated using four different real-world clinical datasets with a total of 26,633 CXRs from 15,097 patients (3,277 COVID-19-positive patients). The area under the receiver operating characteristic curve (AUC) was used to assess diagnostic performance.

**Results**

The AI model trained using a single-source clinical dataset achieved an AUC of 0.82 (95% CI: 0.80, 0.84) when applied to the internal temporal test set. When applied to datasets from two external clinical sites, an AUC of 0.81 (95% CI: 0.80, 0.82) and 0.82 (95% CI: 0.80, 0.84) were achieved. An AUC of 0.79 (95% CI: 0.77, 0.81) was achieved when applied to a multi-institutional COVID-19 dataset collected by the Medical Imaging and Data Resource Center (MIDRC). A power-law dependence, $N^k$ ($k$ is empirically found to be -0.21 to -0.25), indicates a relatively weak performance dependence on the training data sizes.

**Conclusion**

COVID-19 classification AI model trained using well-curated data from a single clinical site is generalizable to external clinical sites without a significant drop in performance.

**Summary**

AI model trained using properly curated, the single-source dataset is generalizable to external sites for the classification of COVID-19 using CXRs, and performance is only weakly dependent on the sample size of the training data.

**Key Points**

- A COVID-19 chest x-ray classification model trained using data from a single clinical site demonstrated generalization to external test cohorts, with an AUC range of 0.79-0.82.
- The model's performance has a weak power-law relationship with the training data size, $N^k$, with the exponent k ranging from -0.21 to -0.25.
- Small training datasets (~100 patients) can be used to develop a baseline AI model with good initial performance, suggesting the importance of data quality over data size in medical AI model development for this application.

**Abbreviations**

AI: artificial intelligence

AUC: area under the receiver operating characteristic curve

COVID-19: coronavirus disease 2019

CXR: chest x-ray radiograph

RT-PCR: reverse transcriptase polymerase chain reaction

**Introduction**

In recent years, deep learning algorithms have shown great promise in medical image analysis to help radiologists and clinicians in disease detection, classification, and severity assessment, thanks to the development of hardware and algorithms in computer vision (1,2). This promise has attracted tremendous research interest over the past two years as the COVID-19 pandemic hit the world and put enormous pressure on healthcare systems. Rapid diagnosis and patient triage play a vital role, especially in the early stage of the pandemic and in resource-limited settings. In response to this urgent need, researchers have rushed to develop AI-based diagnostic models using chest x-rays (CXRs), as shown by the massive uptake in the number of publications on this subject (3). While hundreds of AI models for COVID-19 diagnosis using CXRs have been developed and claimed excellent performance, most of the models failed in generalization when tested externally, as identified in several systematic reviews (4–6).

The poor generalizability of AI models is often attributed to the size and quality of the training data. Without proper data collection and curation strategies, spurious confounding factors, i.e., shortcuts, may exist in the training data (7,8). The model will learn these shortcuts rather than the desired disease features. When shortcuts exist in the training dataset, the trained model can achieve extremely high performance when tested on the internal test set consisting of identically and independently distributed samples from the training distribution. However, the performance may degrade to the chance level when tested externally on real-world clinical datasets.

It is often assumed that increasing the data size and collecting data from diversified sources, i.e., multiple institutions will ensure generalizable models (9). However, such strategies may be difficult to implement due to the regulatory challenges in medical data collection and sharing practices. Furthermore, these strategies may become sub-optimal in a pandemic where solutions must be generated quickly and reliably (10).

This study addresses two fundamental questions in medical AI model development: 1) Is it possible to develop generalizable AI models from a carefully curated single-source dataset? Namely, can a COVID-19 classification model trained from a carefully curated single-source dataset be generalized to external sites? 2) How does the performance of a generalizable AI model depend on the sample size of the well-curated training dataset?

## Materials and Methods

The Institutional Review Boards at Henry Ford Health (Detroit, MI) and the University of Wisconsin-Madison (Madison, WI) approved this retrospective study. Written informed consent was waived because of the retrospective nature of the data collection and the use of fully de-identified images.

**Dataset for model development**

For model development, we collected CXRs from patients with confirmed COVID-19 diagnoses from Henry Ford Health, which includes five hospitals. Patients that underwent frontal view CXR had COVID-19 diagnosis confirmed by RT-PCR test and performed imaging between March 1, 2020, and September 30, 2020, were included. Both COVID-19-positive and COVID-19-negative data are collected within the same time frame. This ensures that the data distribution reflects the patient cohorts in which the algorithm is deployed and removes potential risks of shortcut learning when using pre-pandemic data as the control group. Patients under the age of 18 were excluded.

To ensure the imaging data were consistent with the RT-PCR test, CXRs were excluded if performed more than seven days before or after the RT-PCR test, i.e., a delta window of [-7,7] days. It is well known that some COVID-19 patients may be asymptomatic or minimally symptomatic without findings of pulmonary infection on CXR (11). These cases were excluded for model training and performance evaluation since no disease information was encoded. An independently trained abnormality detection model was used in this study to remove these cases. The details of the abnormality detection model are provided in **Appendix A1**. After applying the abnormality filter, thedataset for model development (HF-Train) consists of 6,689 COVID-19-positive CXRs from 3,264 patients and 10,848 COVID-19-negative CXRs from 4,802 patients. Patient demographics are provided in **Table 1**.

**Datasets for generalizability evaluation**

**Temporal internal test set.** It is important to test model performance in prospectively collected datasets. Namely, the dataset curated in a time frame follows the training data collection time frame. This is referred to as an internal temporal test set. For this purpose, a test set obtained from Henry Ford Health was used for the temporal evaluation of the trained model. This test set consists of consecutive patient cases received for the month of October 2020. A narrow delta window of [-3,3] days was used to ensure testing integrity by only including CXRs performed close to the RT-PCR test. After applying the same abnormality filter to the dataset, the resulting dataset (HF-temporal) consisted of 466 COVID-19-positive CXRs from 334 patients and 5,224 COVID-19-negative CXRs from 3,120 patients.

**BIMCV External test set.** BIMCV is a public COVID-19 chest x-ray dataset collected from 11 hospitals in the Valencian Region, Spain, between February and April 2020 (12). A narrow delta window of [-3,3] was used to include CXRs performed close to the RT-PCR test. After applying the same abnormality filter to the dataset, the resulting dataset consisted of 3,144 COVID-19-positive CXRs from 2,004 patients and 3,335 COVID-19-negative CXRs from 2,365 patients. The list of patient cases used for evaluation can be found on the GitHub repository: https://github.com/uw-ctgroup/CV19-Net-V2.

**UW Health External test set.** This dataset contains consecutive patient cases from the University of Wisconsin Hospitals and Clinics (UW Health), collected from March 2020 to September 2021, with the same inclusion criteria used above. The dataset comprised 694 COVID-19-positive CXRs from 425 patients and 5,435 COVID-19-negative CXRs from 3,574 patients.

**MIDRC External test set.** The Medical Imaging & Data Resource Center (MIDRC) provides an open, diverse, and multi-institutional COVID-19 imaging dataset. Using similar inclusion criteria, a total of 1,022 COVID-19-positive CXRs from 514 patients and 7,313 COVID-19-negative CXRs from 2,761 patients were included in this study. The python script used to download and curate the data cohort can be found on the GitHub repository: https://github.com/uw-ctgroup/CV19-Net-V2.

**Figure 1** shows the image intensity distribution (histogram) of CXRs from each test sites.

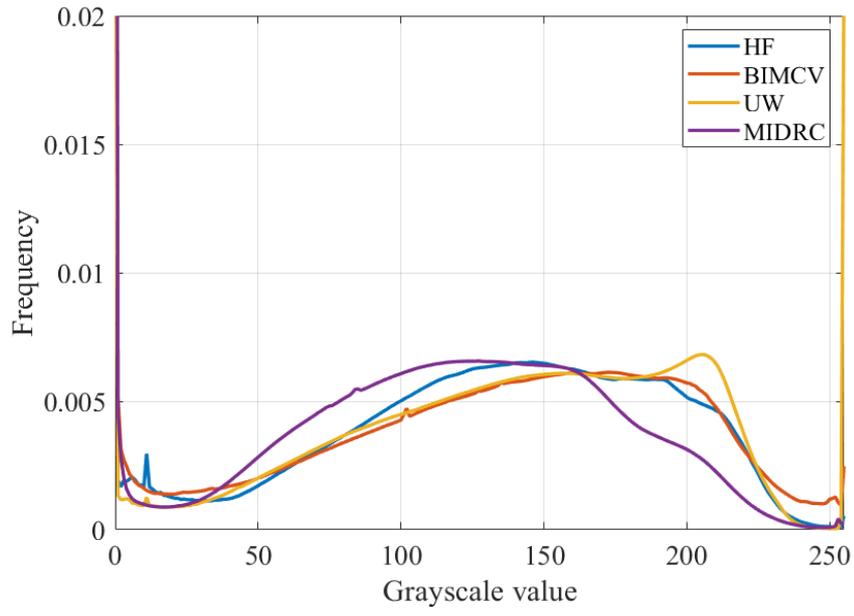

**Figure 1. image intensity distribution (histogram) of CXRs from each test sites.**

**Table 1. Characteristics of CXR datasets used for model training and evaluation**

| | Model development | Clinical test sets | | | |
|---|---|---|---|---|---|
| | HF | HF temporal | BIMCV | UW Health | MIDRC |
| No. images (+/-) | 6689/10848 | 466/5224 | 3144/3335 | 694/5435 | 1022/7313 |
| No. patients (+/-) | 3264/4802 | 334/3120 | 2004/2365 | 425/3574 | 514/2761 |
| Men (+/-) | 1605/2549 | 190/1632 | 1164/1303 | 259/2033 | 281/1230 |
| Age (+/-) | 63±17/69±15 | 66±16/67±18 | 67±17/69±20 | 60±17/64±18 | 59±17/62±16 |
| Imaging system vendors | Carestream (53%), Konica Minolta (20%), GE (19%), Agfa (6%), others (2%) | Carestream (54%), Konica Minolta (16%), Agfa (16%), GE (10%), others (4%) | Agfa (37%), Carestream (13%), Konica Minolta (12%), Philips (8%), Siemens (6%), Canon (6%), others (18%) | Philips (85%), FUJIFILM (8%), others (7%) | Not Available |

**Sampled training datasets with different sizes**

To investigate the impact of training data size on the performance and generalization of the model. Models were developed using training datasets with CXRs from 100, 200, 400, 800, 1200, 1600, 2000, and 6000 patients (some patients may have more than one CXR) sampled from the entire training dataset with a class ratio of 1:1. For each size, ten random samplings were performed. The mean and the standard deviation of the AUC were evaluated. The fitting of AUC as a function of data size was performed using the nonlinear regression function (nlinfit) in MATLAB (version 2022a). Confidence intervals are calculated using the nlpredci function.

**Image preprocessing and model training**

The Digital Imaging and Communications in Medicine (DICOM) files of the collected CXRs were resized to 224 x 224 pixels and saved as 8-bit Portable Network Graphics grayscale images. The neural network model used in this work is an ensemble of five individually trained DenseNet-121 models (13,14) with different training-validation splits. The model was pre-trained using the ImageNet dataset and the ChestX-ray14 dataset (15) from NIH, which consists of 112,120 frontal-view X-ray images of 30,805 unique patients with 14 common disease labels. Details of the model training are provided in **Appendix A2**.

**Statistical analysis**

To evaluate the diagnostic performance of the trained model, the area under the receiver-operating-characteristic curve (AUC), sensitivity, and specificity were calculated. The 95% confidence intervals (CI) for AUC, sensitivity and specificity were calculated using the statistical software R (version 4.0.0) with the pROC package (16). CIs were calculated using the bootstrap method with 2000 bootstrap replicates.

**Model Availability**

The model and datasets used in model training and evaluation will be publicly available for the community to evaluate new algorithms.

## Results

**Clinical test performance of the model**

As shown in **Table 2**, the performance of the trained model on the external test sets is consistent with internal temporal validation, even though the test population and imaging system vendor distributions are entirely different between the training and the external test sets. For sensitivity and specificity calculations, no dataset-specific threshold tuning was applied; a fixed threshold value of 0.7 was used for all tests, further demonstrating the generalizability of the trained model.

Table 2. Clinical test performance of the model

|             | HF temporal       | BIMCV             | UW Health         | MIDRC             |
|-------------|-------------------|-------------------|-------------------|-------------------|
| AUC         | 0.82[0.80,0.84]   | 0.81[0.80,0.82]   | 0.82[0.80,0.84]   | 0.79[0.77,0.81]   |
| Sensitivity | 0.51[0.46,0.55]   | 0.52[0.50,0.54]   | 0.50[0.46,0.53]   | 0.45[0.41,0.47]   |
| Specificity | 0.94[0.93,0.95]   | 0.91[0.90,0.92]   | 0.95[0.94,0.96]   | 0.92[0.91,0.93]   |

Note: A fixed threshold value of 0.7 was applied for sensitivity and specificity calculations. Numbers in the square brackets represent a 95% confidence interval (CI).

**Performance of the model as a function of training data size**

**Figure 2** shows the AUC of the model on the four test sets as a function of the training data size. The AUC values corresponding to the data size of 100, 200, 400, 800, 1200, 1600, and 2000 are used to fit the parameters of a learning curve, in the form of $y = aN^k + b$, where $y$ is the AUC of the model, $N$ is the training data size (number of patients), $a, k,$ and $b$ are parameters. Incredibly, the AUC predicted from this fitted function for the data size = 6000 almost precisely matches the actual measured AUC, showing excellent predictive power of the fitted function at larger sample sizes. The performance of models with different training data sizes is also reported in **Table 3**. As seen from these results, a significant performance gain can be observed when the raining data size increases from 100 patients to 800 patients. While adding more training data would continually improve the model's performance, the performance gain rapidly becomes marginal. This weak power-law relationship is consistent with that reported in (17) for several computer vision applications, where they show typical $k$ values ranging from -0.07 to -0.35.

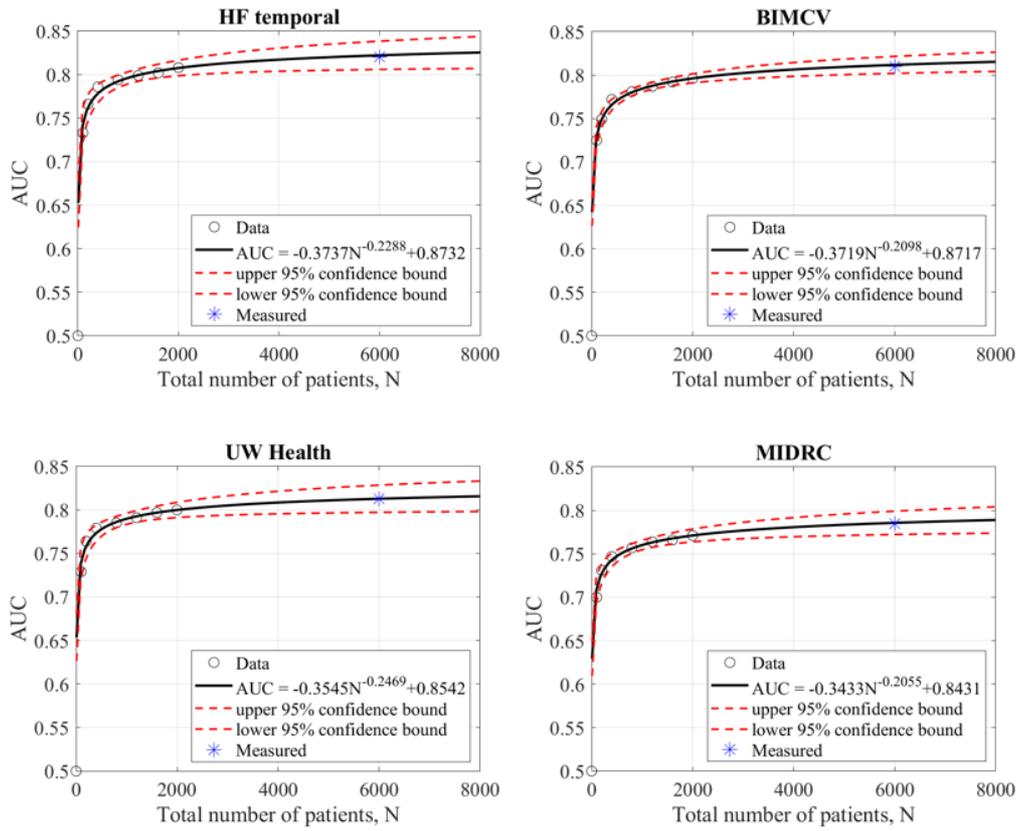

**Figure 2.** AUC vs. training data size. Note that a data point with N=1 and AUC = 0.5 was manually added for all plots.

Table 3. AUC vs. training data size

|  | HF temporal | BIMCV | UW Health | MIDRC |
|---|---|---|---|---|
| 100 | 0.732 ±0.016 | 0.725 ±0.023 | 0.729 ±0.024 | 0.700 ±0.025 |
| 200 | 0.766 ±0.026 | 0.749 ±0.028 | 0.764 ±0.018 | 0.731 ±0.019 |
| 400 | 0.786 ±0.009 | 0.772 ±0.015 | 0.779 ±0.009 | 0.747 ±0.010 |
| 800 | 0.794 ±0.007 | 0.781 ±0.009 | 0.785 ±0.006 | 0.757 ±0.009 |
| 1200 | 0.799 ±0.008 | 0.787 ±0.007 | 0.791 ±0.005 | 0.764 ±0.009 |
| 1600 | 0.802 ±0.003 | 0.792 ±0.005 | 0.797 ±0.006 | 0.766 ±0.005 |
| 2000 | 0.808 ±0.004 | 0.796 ±0.005 | 0.800 ±0.005 | 0.771 ±0.006 |
| 6000 (prediction) | 0.819 [0.805 0.834] | 0.811 [0.802 0.820] | 0.813 [0.797 0.829] | 0.786 [0.772 0.799] |
| 6000 (measured) | 0.818 | 0.809 | 0.813 | 0.786 |

**Notes: The number after ± represents the standard deviation of ten measurements. Data in the square brackets are 95% confidence intervals (CIs) of the prediction.**

**Discussion**

To the best of the authors' knowledge, this is the largest multi-site external evaluation of an AI model for the classification of COVID-19 using chest CXRs, with more than 15,000 patients and 25,000 CXRs. Previous publications on COVID-19 classification using CXRs have either only been internally tested or tested on a single external test set within a short time range and with much fewer cases (18–20). The homogeneous performance of our model on multiple external test datasets may be partially attributed to the quality of the training dataset. We did not use pre-pandemic controls in model training and testing. Both training and testing focus on the target population cohort with suspected COVID-19.

While the clinical value of developing AI algorithms for the detection of COVID-19 using chest radiographs may not be of primary clinical interest in this stage of the pandemic, given that RT-PCR and rapid antigen tests are now widely available, however, the clinical significance of this work can be summarized as follows.

First, despite hundreds of models being developed two years into the pandemic, it remains unclear what is the real-world performance of AI models in the classification of COVID-19 using chest radiographs due to the lack of external testing with real-world clinical datasets. This work, together with another recently published work (21), provides a reference for AI's performance on real-world datasets. The reported sensitivity and specificity of the AI model in this work are similar to those of radiologists, as shown in a recent study by (22).

Second, while it is a common belief that the use of data from diversified sources is necessary for generalizable AI models (9), in this work, we have shown that with careful data curation, it is possible to train AI models with good generalizability using data from a single source. One potential benefit of single-source data is that the risk of spurious correlation between the data source and the disease label can be well controlled during the data curation process. This contrasts with multi-center data collection, which requires tremendous resource allocation nationally and internationally to reduce the risks of introducing shortcuts (i.e., the spurious correlations between the medically irrelevant features with the training labels) in the collected datasets.

Third, this is the first time shown that an AI model trained using a well-curated single-source dataset (HF-train) is also generalizable for another clinical site (UW Health) and other publicly available clinical datasets (BIMCV and MIDRC) without significant performance degradations.

Finally, a weak power-law relationship was found between the model's generalization performance and the training data size. The results show that even a small data size, such as 100 patients, can provide a good baseline model with high-quality data. This baseline model can provide initial clinical value when deployed in the clinical workflow. Starting with a baseline model, the model can be continuously updated to improve its performance in a real-world clinical setting.

This study also has several limitations. First, the real-world performance of AI models to detect and triage suspected COVID patients can only be rigorously evaluated with a prospective study. Namely, the model should only be applied to suspected patients who do not have a confirmed viral test result at the imaging time. Although we attempted to simulate a prospective patient cohort by limiting imaging studies performed within three days before or after the RT-PCR test, the patient cohort may still be slightly different from the prospective cohort. Second, the model's performance was not compared to that of the radiologists. This is because diagnosing COVID-19 from chest radiographs was not a routine clinical task for radiologists. Therefore, the comparison does not carry too much clinical significance. Third, based on the model's performance, AI models are insufficient to be used as a stand-alone diagnostic tool for COVID-19; their clinical value is more along with the optimization of clinical workflow for radiologists by prioritizing the reading lists in the clinic. However, evaluating their value requires deploying models in the actual clinical workflow. The generalizability of AI models must be proven before clinical deployment can be initiated, as was done in this work. Another potential limitation is that the virulence of the SARS-CoV2 virus has evolved during the course of the pandemic, and that the majority of the population has acquired some level of immunity due to vaccination or prior infection. While beyond the scope of this work, further investigation into the evolving nature of COVID-19 pneumonia and its impact on the AI-model in this work is needed.

In conclusion, a COVID-19 classification model trained using properly curated data from a single clinical site was evaluated on multiple external test sites from different regions and countries and with varying care settings. The model demonstrated consistent performance across all sites and thus is generalizable.

**Acknowledgments**

This study was partially supported by NIH Grants 3U01EB021183-W4, R01HL153594, R01EB032474, and a grant from the Wisconsin Partnership Program.

Dr. Reeder is a Romnes Faculty Fellow, and has received an award provided by the University of Wisconsin-Madison Office of the Vice Chancellor for Research and Graduate Education with funding from the Wisconsin Alumni Research Foundation.

The imaging and associated clinical data downloaded from MIDRC (The Medical Imaging Data Resource Center) and used for research in this [publication/press release] were made possible by the National Institute of Biomedical Imaging and Bioengineering (NIBIB) of the National Institutes of Health under contracts 75N92020C00008 and 75N92020C00021. The content is solely the responsibility of the authors and does not necessarily represent the official views of the National Institutes of Health.

**Appendix A1: Abnormality detection model**

A deep neural network classifier was trained to identify chest x-ray radiographs (CXRs) without pneumonia or lung opacity. To train the classifier, 94,967 CXRs were selected from the MIMIC-CXR dataset (1). Images with positive lung opacity, lung lesion, or pneumonia labels were assigned to the positive class. In contrast, those with negative labels for the above disease labels and those with "No Finding" labels were assigned to the negative class. Images were randomly shuffled and partitioned into 80% training and 20% validation. Image partitions were based on the per-patient strategy to avoid data leakage.

The classifier model was developed using the DenseNet-121 (2) architecture with weights pre-trained from the ImageNet dataset (3). DICOM images were resized to $224 \times 224$ using bilinear interpolation and converted to 8-bit images as model inputs. Adam optimizer with an initial learning rate of $5 \times 10^{-5}$ was used to minimize the Binary Cross-Entropy loss. Image augmentations were applied, including random horizontal flipping and rotation (between 0-30 degrees). If validation loss did not improve over five consecutive epochs, early stopping was used.

The trained model achieved an AUC of 0.86 [0.85, 0.87] on the validation set. Using a threshold of 0.2 over the output probability scores, the sensitivity and specificity were determined to be 0.90 and 0.67, respectively. This threshold was also used to remove the cases without pneumonia or lung opacity (abnormality score less than 0.2) in the COVID-19 chest x-ray datasets used in this work.

**Appendix A2: Technical details of model training**

The deep neural network model for COVID-19 pneumonia classification was developed using the DenseNet-121 (2) feature extraction backbone. The final classifier was modified to a two-class output with a Softmax activation function. DICOM images were resized to $224 \times 224$ using bilinear interpolation and converted to 8-bit grayscale images as model inputs. The image intensity value mapping was based on the window level and window width attributes in the DICOM file. Adam optimizer with an initial learning rate of $5 \times 10^{-5}$ was used to minimize the binary Cross-Entropy loss. Image augmentations were applied, including random horizontal flipping, rotation (between 0-30 degrees), brightness, and contrast. Early stopping was executed if validation loss did not improve over five consecutive epochs.

The dataset for model development was split into 80% training and 20% validation per patient level. Five models were trained with identical hyperparameters but different random seeds in model initialization and different randomly sampled validation sets. A quadratic mean of the prediction probability scores was taken to generate the final prediction probability score: $S = \left[\frac{1}{5}\sum_{i=1}^{5} S^2(i)\right]^{1/2}$.